# Efficient Thermal Transport across Molecular Chains in Hybrid 2D Lead Bromide Perovskites

Nabeel S. Dahod[1], Watcharaphol Paritmongkol[1,2], William A. Tisdale[1]*

[1]Department of Chemical Engineering, Massachusetts Institute of Technology, Cambridge, Massachusetts, 02139, USA.

[2]Department of Chemistry, Massachusetts Institute of Technology, Cambridge, Massachusetts 02139, USA.

*Correspondence to: tisdale@mit.edu

**Abstract:** We report measurements of the heat capacity and cross-plane thermal conductivity of 2D $(C_xH_{2x+1}NH_3)_2[MAPbBr_3]_{n-1}PbBr_4$ (MA = methylammonium) lead bromide perovskites (2D LHPs) at room temperature as a function of both the octahedral layer thickness (n = 1,2,3) and the organic spacer chain length (x=4,5,6,7,8) using differential scanning calorimetry (DSC) and frequency domain thermoreflectance (FDTR) respectively. We observe ultralow thermal conductivities (0.18 – 0.51 W/m·K) for all 2D LHPs studied, but surprisingly minimal suppression of thermal conductivity with respect to bulk $MAPbBr_3$ (0.5 W/m·K). Cross-plane thermal conductivity is found to increase monotonically as a function of both the octahedral layer thickness (0.18-0.26 W/m·K for n=1-3) and the organic chain length (0.18-0.51 W/m·K for x=4-8). Additionally, we measure heat capacities that are well described by composite theory, suggesting bulk-like phonon density-of-states within the separate organic and inorganic subphases of the layered structure. The striking observation of increasing thermal conductivity with increasing organic phase fraction (i.e. increasing organic chain length) indicates efficient thermal transport along the ordered alkyl chain backbone. Our experimental results agree most closely with a predictive model of ballistic phonon transport with diffuse interface scattering – rather than normal thermal conduction within each phase. This study indicates the potential for synthesizing 2D LHPs with thermal conductivity that exceeds the bulk perovskite phase, while also shedding light on relevant phonon transport pathways in 2D LHPs.

**Introduction:**

Hybrid organic-inorganic perovskites (HOIPs) have emerged over the past decade as a potential candidate to supplant conventional semiconductors for use in a variety of device applications, including efficiency-competitive solar cells[1–3], color tunable light emitting diodes (LEDs)[4,5], sensitive photodetectors[6], and lasers[7]. These materials are denoted via the chemical formula $ABX_3$, and exhibit a crystal structure in which a small organic cation (A), such as methylammonium (MA), is enclosed within metal (B) halide (X) octahedra. HOIPs exhibit high optical absorption coefficients, small exciton binding energies, and long/balanced electron-hole diffusion lengths[8,9]. In addition to these 3D semiconductors, a variety of nanostructured HOIPs have been developed as alternative material architectures. One increasingly relevant nanostructured perovskite is the 2D Ruddlesden-Popper organic-inorganic lead halide perovskite (2D LHP), in which atomically thin sheets of the traditional perovskite structure are separated by bilayers of larger organic cations.[10] 2D LHPs have shown improved stability and promising performance metrics in solar cells and LEDs in particular.[10–16] These periodic layered nanostructures can be grown as macroscopically large crystals, with long-range order and minimal grain boundaries. Additionally, since the perovskite sheets are thin enough to exhibit quantum confinement, the size of this phase can be controlled independently for further tunability of the optical and electronic properties.[17,18] Finally, the organic subphase can be manipulated to potentially improve performance, particularly in terms of the exciton binding energy and photoluminescence quantum yield.[19–21]

In addition to optical and electronic properties, thermal and vibrational phenomena in bulk (3D) and 2D LHPs are of emerging interest[22–36]. Thermal management is of critical concern in the development of any material for widespread use in optoelectronic device applications, particularly when heating can degrade material performance.[37,38] In addition to mitigating device-heating, engineering electron-crystal phonon-glass materials that dissipate heat slowly while efficiently conducting charge opens the door for further applications such as thermoelectric devices.[39]

Unlike conventional semiconductors, bulk HOIPs are characterized by ultralow thermal conductivities (0.34-0.73 W/mK), which are up to two orders of magnitude lower than those for traditional inorganic semiconductors.[22,23,29,31] Unlike metals and highly-doped semiconductors, where charge carriers contribute significantly to overall thermal transport, phonons are believed to be exclusively responsible for thermal transport in HOIPs.[22] From kinetic theory, the phonon thermal conductivity can be expressed in terms of the product of the mean-free path ($\Lambda$, MFP), group velocity ($v$), and heat capacity (C) of the phonons responsible for heat transport as $k = \frac{1}{3}Cv\Lambda$. Preliminary reports have identified that the ultralow measured thermal conductivities likely stem from a combination of factors such as low phonon group velocities and high anharmonic scattering between phonon modes associated with the lead-halide lattice and the small organic cation enclosed within it.[23,30,32–35] The elastic softness of HOIPs corresponds to lower sound speeds of acoustic phonons, which dominate the thermal conductivity in conventional semiconductors.[40] Additionally, the presence of a high density of low frequency optical phonons that couple anharmonically with these acoustic modes via multiphonon scattering processes leads to very short estimated MFPs in HOIP solids (~1 nm).[30] Other studies, however, suggest that the MFP of phonons may be much longer and contend that the phonon speeds are exclusively responsible for the ultralow thermal conductivities of HOIPs.[23]

Although 2D LHPs are derived from their bulk counterparts, their natural structure more closely resembles that of inorganic superlattices, or multiple quantum wells, which are multilayer periodic stacks of ultrathin bulk semiconductor films. In many ways, 2D LHPs present as idealized superlattices, with atomically abrupt interfaces between layers, minimal thickness variation throughout the structure, and exceptional long-range order. In contrast, unlike all-inorganic superlattices, 2D LHPs are composed of phases with a large acoustic impedance mismatch and which are bound to one another by weaker hydrogen-bonding interactions. It should be noted, however, that the impedance mismatch in 2D LHPs is less extreme than in other organic-inorganic hybrids, due to the relative softness of the perovskite subphase. Inorganic superlattices showcase a variety of unique phenomena related to phonon transport, such as extreme suppression of thermal conductivity due to the added thermal boundary resistance between each film in addition to wave-like phonon transport in which collective vibrations of the entire superlattice structure are largely responsible for thermal energy transport.[41–46] In the latter scenario, the wave nature of bulk phonons is maintained and the material can be treated as its own bulk solid with distinct diffusive phonons, whereas for the former the transport phenomena is as described by ballistic (minimally scattered) phonon propagation within each sub-layer that scatter exclusively at the boundaries between layers.[45,47]

Guo *et al.* probed the existence and coherence time of coherent longitudinal acoustic phonons of the entire 2D LHP multi-layer structure, determining that for MAPbI-based 2D LHPs such modes exist but travel much slower and with far shorter coherence times than observed for the bulk perovskite.[48] Similarly, Raman signatures that may be assigned to longitudinal acoustic phonons have been identified at frequencies below 30 cm$^{-1}$.[24,27,28] These results suggest the possibility that superlattice modes may contribute to thermal conductivity in much the same way as observed in inorganic superlattices. However, significant anharmonic interaction between the organic and inorganic subphases could diminish the importance of this contribution, a crossover well established for a variety of inorganic superlattices.[44,47,49,50] Giri *et al.* measured thermal conductivity in selected 2D iodide and bromide perovskites, finding thermal conductivities in the range 0.10-0.19 W/m·K, which is lower than the corresponding value in bulk MAPbBr$_3$ (0.51 ± 0.12 W/m·K) or MAPbI$_3$ (0.34 ± 0.08 W/m·K)[23]. In 2D alkylammonium lead iodide perovskites, specifically, cross-plane thermal conductivity has been studied as a function of octahedral layer thickness[25] and alkyl chain length[26]. In both studies, 2D lead iodide perovskites exhibited ultralow thermal conductivities (0.06-0.37 W/m·K) that were, in some cases, heavily suppressed relative to the corresponding bulk lead iodide perovskite.

Here, we investigate thermal transport in a series of systematically varied 2D alkylammonium lead bromide perovskite crystals. We report measurements of the heat capacity and thermal conductivity of 2D (C$_x$H$_{2x+1}$NH$_3$)$_2$[MAPbBr$_3$]$_{n-1}$PbBr$_4$ perovskites at room temperature as a function of both the octahedral layer thickness of the perovskite layer (n = 1,2,3) and the organic spacer chain length (x=4,5,6,7,8) using differential scanning calorimetry (DSC) and frequency domain thermoreflectance (FDTR), respectively. We observe ultralow thermal conductivities (0.18 – 0.51 W/mK) for all 2D LHPs studied, that in some cases were only minimally suppressed relative to the bulk. Significantly, the thermal conductivity was found to increase monotonically as a function of both the octahedral layer thickness and the organic chain length. We find good agreement between experimental observations and a predictive model for ballistic phonon transport with diffuse interface scattering[41]. We discuss the implications of these findings for tuning composite thermal conductivity in 2D LHPs and contrast our observations to other reports.

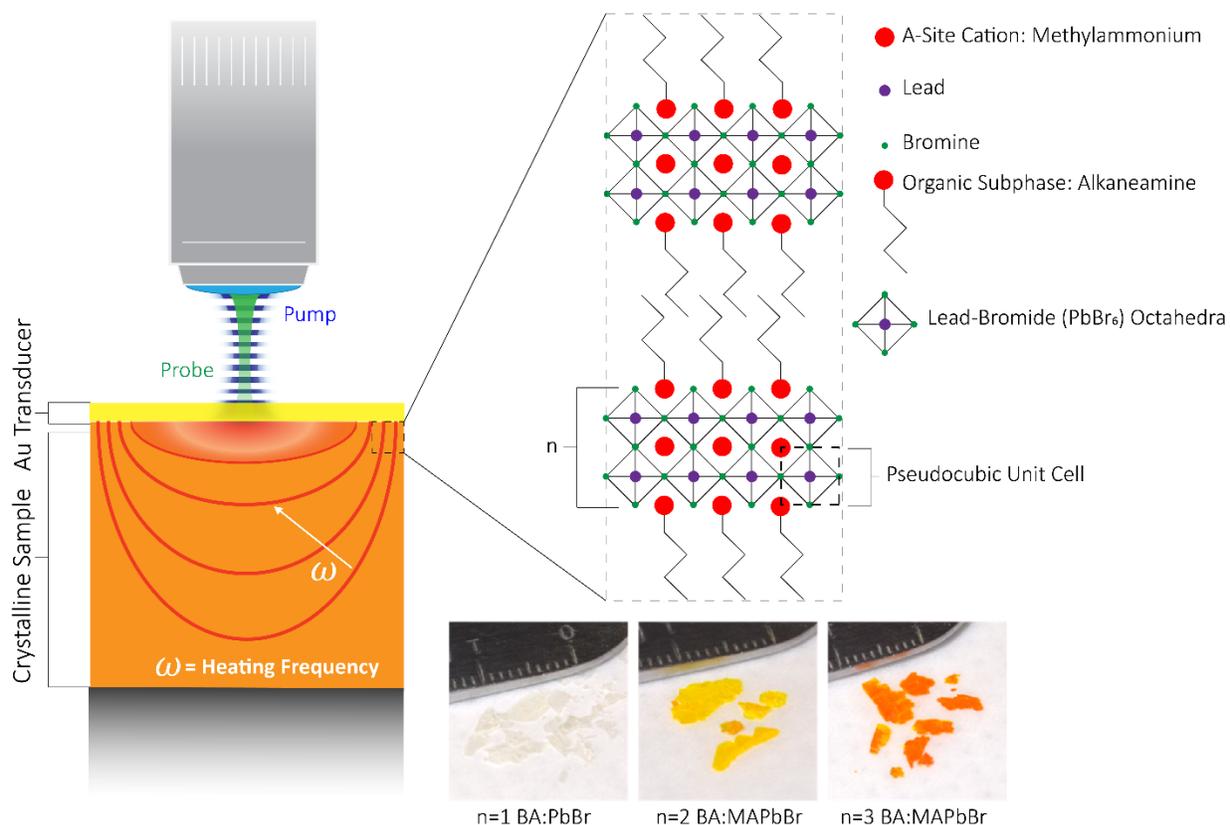

**Figure 1.** *Left*: Schematic of Frequency Domain Thermoreflectance (FDTR) measurement, in which two coaxially aligned lasers are focused through a microscope objective and reflected off of a gold (Au) thin film transducer, which has been exogenously deposited atop a solid sample of interest (here small 2D LHP crystals). The pump laser sinusoidally heats the transducer and the underlying sample, and the corresponding modulations in the reflected probe intensity inform on heat transport in the sample. The pump modulation frequency, ω, directly determines the penetration depth ($\delta$) of the heating in the measurement according to the scaling relationship $\delta \sim \sqrt{\alpha/\omega}$ where $\alpha$ is the thermal diffusivity of the solid. The amplitude and phase shift of the reflected probe signal are recorded as a function of modulation frequency and fit to a continuum model to extract the effective thermal conductivity of the sample. See Supporting Information for further details. *Right (top)*: Schematic of a typical n=2 2D LHP. The octahedral layer thickness, n, corresponds to the number of pseudocubic unit cells within the inorganic subphase. *Right (bottom)*: Photographs of millimeter-sized n = 1,2,3 PbBr 2D LHP solid crystals.

**Results & Discussion:**

*Frequency Domain Thermoreflectance (FDTR)*

FDTR is a widely utilized optical metrology technique for monitoring thermal transport in solids, small crystals, thin films, and across molecular junctions such as self-assembled monolayers.[51–54]

We implemented a home-built FDTR system (Figure 1; see Supporting Information for further details) to analyze macroscopic 2D LHP crystals synthesized using a cooling-induced crystallization process.[55] All crystals were adhered to a glass substrate and coated with a thermally evaporated Au film (100 nm). Two coaxially aligned continuous-wave lasers were focused on the sample of interest through an objective lens, one of which ("pump", λ=488nm) was intensity modulated at kHz to MHz frequency while the other ("probe", λ=532nm) was not. The Au film transducer exhibits linear thermoreflectance at the laser wavelengths/powers employed, leading to modulation of the reflected probe beam intensity at the pump heating frequency, but phase-shifted by an amount that depends on the thermal transport characteristics of the underlying perovskite sample. The reflected probe light is separated from the pump and its phase lag across a range of pump frequencies is catalogued using a lock-in detection scheme. This frequency response is fit to a continuum model of heat transport based on the diffusion equation for the system geometry in order to extract the effective thermal conductivity of the sample.[56] Since we operate using heater spot sizes (~2 μm FWHM) larger than the penetration depths of the measurement (~200-1000 nm), this extracted value specifically corresponds to an effective cross-plane thermal conductivity of the macroscopic crystal in the surface-normal direction. In order to obtain reliable measurements of the thermal conductivity, the transducer thickness, laser spot size, and heat capacity are measured separately using profilometry, a CCD camera, and DSC respectively. Further details on the FDTR technique, sample preparation, and crystal synthesis are provided in the Supplementary Information.

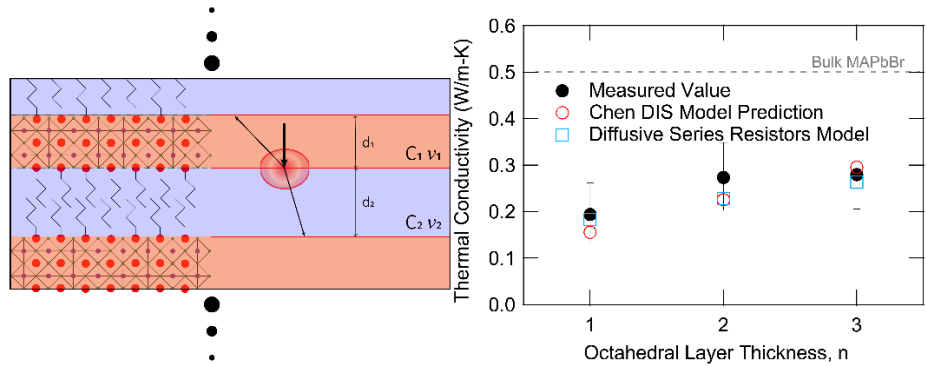

**Figure 2.** *Left*: Schematic of 2D LHP and approximation of it as a periodic multilayered structure. Parameters used to approximate each layer are shown within their respective shaded region, and correspond to the sub-layer thickness (d), volumetric heat capacity (C) and sound velocity (*v*). Black arrows show phonon transport as represented by model in Eq. 1. Phonons move ballistically through each layer and scatter diffusely (annihilatively) at the interfaces. *Right*: Measured thermal conductivity using FDTR for n = 1 BAPbBr and n = 2,3 BA:MAPbBr 2D LHPs at room temperature (black) and comparison to the theoretical model in Eq. 1 (red).

*Dependence on Octahedral Layer Thickness*

At room temperature, the experimentally measured thermal conductivity of PbBr 2D LHPs containing butylamine as the organic spacer (n=1 C4:PbBr and n=2,3 C4:MAPbBr) increases monotonically with octahedral layer thickness from $k = 0.19 \pm 0.07$ for n=1 to $k = 0.28 \pm 0.08$ W/m·K for n=3 (Figure 2). These values represent an approximately 50% reduction in the thermal conductivity compared to the bulk MAPbBr$_3$ value, which was measured to be $k = 0.49 \pm 0.1$ W/m·K on the same apparatus – consistent with previous reports.[23] While suppression of phonon transport is well documented in nanostructured semiconductors[57], the magnitude of the reduction that we observe is surprisingly small. For example, colloidal semiconductor nanocrystal solids show a thermal conductivity reduction of two orders of magnitude relative to their bulk values, due largely to impedance mismatch at the organic-inorganic interfaces between the semiconductor crystallites and their surface-bound molecular ligands.[53] For the 2D LHPs studied here, even if no reduction in the thermal conductivity within each subphase is assumed and a simple series-resistance model ($\frac{L_{inorganic}}{k_{inorganic}} + \frac{L_{organic}}{k_{organic}} = \frac{L_{2DLHP}}{k_{2DLHP}}$) is used to estimate the "upper limit" of the thermal conductivity for the 2D LHP in the diffusive transport regime (plotted in orange in Figure 2), the low thermal conductivity of the organic subphase (~0.1 W/mK for most short-chain alkanes in their pure phase) actually predicts a lower thermal conductivity than measured (~0.15-0.21 W/mK for n=1-3 C4 PbBr 2D LHPs). This is especially surprising given that this representation does not account for any phonon scattering at the organic-inorganic interfaces between layers.

The relatively modest reduction in thermal conductivity of 2D LHPs relative to their bulk perovskite counterparts suggests that the average MFP of heat carrying phonons in 2D LHP crystals is not appreciably larger than the thickness of the inorganic layers (1-2 nm). This is largely in agreement with several recent computational and experimental studies, and reinforces the prevailing belief that thermal transport in bulk HOIPs is limited by short MFP phonons.[30,32,33] Additionally, and more importantly, since our measured thermal conductivity in 2D LHPs is actually higher than that predicted from traditional composite theory (even when neglecting interface resistances), the dominant phonon scattering mechanism within 2D LHPs is fundamentally different than in the bulk organic and inorganic reference phase.

In order to understand the monotonic increase observed as a function of the octahedral layer thickness, we utilize an expression for the cross-plane thermal conductivity, $k$, of periodic multilayer solids derived from the Boltzmann transport equation (BTE) by Chen *et al.* for use with all-inorganic superlattices (Eq. 1),

$$k = \frac{1}{2}\left(\frac{1}{C_1 v_1} + \frac{1}{C_2 v_2}\right)^{-1} \frac{(d_1 + d_2)}{2}, \qquad (Eq. 1)$$

where the subscripts denote different sub-phases of the layered material, $C_i$ is the volumetric heat capacity, $v_i$ is the sound velocity, and $d_i$ is the layer thickness.[41] This expression is obtained for a periodic structure under the relaxation time approximation, and examines the limiting case in which layers are very thin (with periodicity comparable to the MFP of phonons within either film) and exhibit infinite long-range order with minimal defects or heterogeneities in the layers. As a result, the transport within each layer is ballistic and scattering only occurs at the interfaces, hence the series representation of the resistances $\left(\frac{1}{Cv}\right)$ in each layer. This representation implies fully diffuse inelastic phonon scattering, in which phonons scattered at an interface share no phase relationship with their previous state. Accordingly, this assumption reflects the lowest predicted thermal conductivities for a model invoking ballistic transport and appropriately fits inorganic superlattices in which diffuse scattering is dominant at interlayer interfaces.[41]

Each of these assumptions, while limiting for most inorganic superlattices, are actually rather appropriate for 2D LHPs as described above. The assumption of ballistic transport within each perovskite layer is reasonable given that the octahedral layer thicknesses studied here (<2 nm) are comparable to or less than the phonon MFP of bulk HOIPs. [30,32,33] (Although, it is worth noting that scattering within the perovskite layers – as quantified through the addition of a series resistance $\left(\frac{d_{inorganic}}{k_{MAPbBr,bulk}}\right)$ corresponding to diffusive transport in the octahedral layer – would minimally impact the predicted conductivity.) More significant is the assumption of ballistic transport through the butylammonium organic spacer layer. This assumption calls for replacing the ultralow thermal conductivities of liquid alkanes with significantly faster ballistic thermal transport across the molecular junctions that comprise the organic subphase. Though initially surprising, such behavior has been observed elsewhere as ballistic heat transport across the molecular junctions within self-assembled monolayers[58,59].

We find reasonable agreement between the diffuse inelastic superlattice (DIS) model represented by Eq 1 and our experimental data when using bulk values for the heat capacity and sound speed within the perovskite layers, sound speeds for tethered aliphatic chains extracted from previous studies on SAMs[60] and colloidal nanocrystals[61], and heat capacities for liquid alkanes (which have comparable densities to the organic subphase in 2D LHPs). While it is routine to approximate the phonon group velocity as the sound speed, using the volumetric heat capacity can become tenuous when optical phonons that contribute significantly to the heat capacity do not contribute to thermal transport. This, however, does not seem to be the case for HOIPs since optical phonons contribute heavily to thermal conductivity.[30]

*Heat Capacity of 2D LHPs*

We experimentally measured the heat capacity of the 2D LHP crystals used in this study by differential scanning calorimetry (DSC) and found the values to be consistent with those estimated from mass-weighted averages of the bulk values (Figure 3). This finding further reinforces the view that each nanoscale subphase retains its bulk heat capacity and sound velocity. Similar conclusions have been reached for other hybrid organic-inorganic nanomaterials such as colloidal quantum dot solids, wherein the vibrational density of states (vDOS) of the nanomaterial is simply the sum of the vDOS of the organic (ligand) and inorganic components.[53]

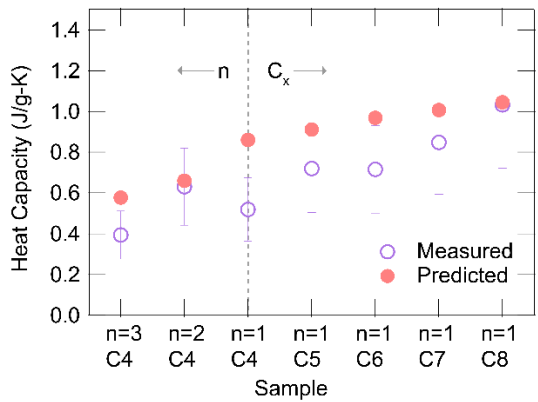

**Figure 3.** Experimentally measured (purple) specific heat of 2D LHPs with various organic spacer molecules (chain length C) and octahedral layer thicknesses (n) using DSC agrees with predictions (red) using a mass-weighted average of the bulk heat capacities ($C_{2DLHP} = fC_{MAPbBr,bulk} + (1-f)C_{Alkane,bulk}$), where $f$ is the mass fraction of the perovskite subphase.

Increasing the octahedral layer thickness of 2D LHPs, using the DIS model, should lead to an increase in the thermal conductivity solely from the increase in the period thickness, as we observe (i.e. there are fewer scattering interfaces per unit length). Within this model framework, the interfaces provide the same thermal resistance regardless of the octahedral layer thickness, and the phonon transport within the perovskite layer is ballistic. This also potentially explains why the measured increase in thermal conductivity moving from n=2 to n=3 for C4:MAPbBr is less than that predicted by the DIS model (Figure 2). As the octahedral layer thickness approaches the phonon MFP, the assumption of purely ballistic transport breaks down. The decrease in measured thermal conductivity for n=3 C4:MAPbBr with respect to theory may be due to phonon-phonon scattering within the perovskite layers.

*Dependence on Organic Chain Length*

To probe the limits of ballistic transport within the organic subphase, we measured room temperature thermal conductivities for n=1 PbBr 2D LHPs with varying organic spacer chain lengths (butylammonium-octylammonium, $C_4$-$C_8$). Strikingly, the cross-plane thermal conductivity of the 2D LHP crystals increased monotonically with increasing organic spacer chain length from 0.19 ± 0.07 W/m·K for n=1 butylammonium:PbBr to 0.51 ± 0.13 W/m·K for n=1 octylammonium:PbBr (Figure 4a). As before, the measured thermal conductivity is well-approximated by the DIS model, within instrumental error. Further analysis of the model predictions reveals that varying the length of the organic carbon chain separating the perovskite layers actually influences the thermal conductivity through multiple channels. Firstly, the period thickness of the organic layer increases with the length of the carbon chain, reducing the spatial frequency of scattering interfaces. Secondly, the sound speed through the bound organic molecules increases by a factor of 3 moving from butylammonium to octylammonium, while the heat capacity decreases only modestly (<10%) over the same range. Thus, the thermal transmittance ($C_2 v_2$, the product of heat capacity and sound velocity) and the period thickness both increase with the length of the organic spacer molecule, without affecting the perovskite layer (Figure 4b).

It is particularly noteworthy that the measured thermal conductivity of n=1 octylammonium:PbBr actually meets that of bulk MAPbBr. While striking, this observation is consistent with expectations from the DIS model. In fact, this behavior highlights an interesting consequence of the DIS model: that the thermal conductivity of 2D LHPs is, in principle, not limited by the same mechanisms that lead to ultralow thermal conductivity of bulk perovskites. The primary driver for this behavior appears to be the ability of the molecular chains to sustain ballistic transport across greater distances than the perovskite phase. As long as the transport within both sub-phases remains ballistic, the only determinants for thermal conductivity are the thermal transmittance ratio $\left(\frac{C_2 v_2}{C_1 v_1}\right)$ and the period thickness. Thus, for relatively efficient ballistic transport (large thermal transmittance) of the organic sub-phase the thermal conductivity will reach a theoretical maximum dependent solely on the period thickness and the thermal transmittance of the perovskite layer. It must be stressed that such a maximum is actually far above what we probe herein, but could significantly exceed the thermal conductivity of bulk MAPbBr.

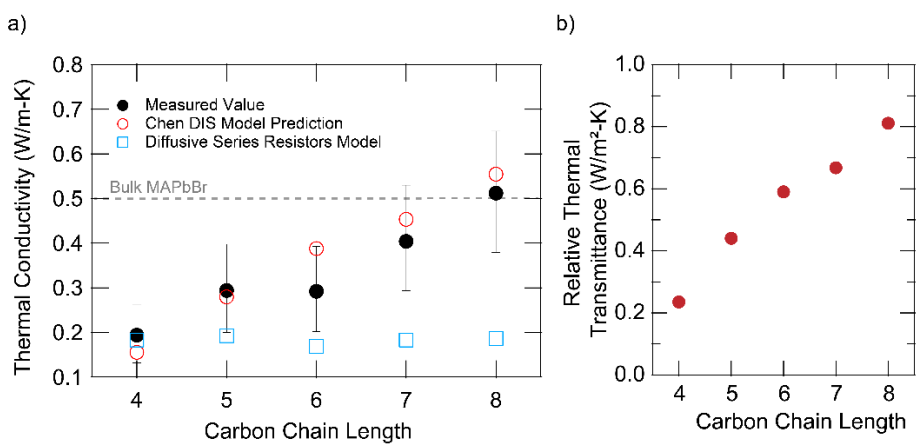

**Figure 4.** a) Measured thermal conductivity using FDTR for n = 1 PbBr 2D LHPs at room temperature (black) and comparison to different theoretical models as a function of the aliphatic hydrocarbon chain length of the organic spacer subphase. b) Calculated relative thermal transmittance ratio $\left(\frac{C_2 v_2}{C_1 v_1}\right)$ for the organic spacers used in (a) based on bulk heat capacities and sound speeds measured from SAMs & alkyl ligands bound to quantum dot surfaces.

Cross-plane thermal conductivity in other 2D LHPs perovskites was studied recently using FDTR[25] and the related technique, time-domain thermoreflectance (TDTR)[26]. Similar to the 2D lead bromide results presented here, the 2D lead iodides also exhibited ultralow thermal conductivity (0.06-0.37 W/m·K) that was suppressed relative to the corresponding bulk value. However, unlike the 2D lead bromide results presented here, the 2D lead iodides exhibited *decreasing* thermal conductivity with increasing octahedral layer thickness[25] and increasing organic chain length[26]. It is unclear whether the opposite trends observed for bromides and iodides arises from true chemical differences in intrinsic material behavior, or from extrinsic effects such as sample quality or measurement methodology. (We note, however, that the thermal conductivities of bulk $MAPbBr_3$ and BA$_2$PbBr$_4$ measured on our instrument are identical, within error, to previous reports[23,27] – suggesting that measurement methodology is not the origin of different trends.) Unfortunately, attempts to measure the thermal conductivity of 2D lead iodide perovskites using our FDTR instrument were unsuccessful. 2D lead iodide samples readily degraded during FDTR measurement in our lab, presumably due to strong resonant excitation of the iodide band gap by the 488 nm pump or 532 nm probe lasers – which is not a problem for the blue/UV-absorbing bromide perovskites.

Table 1 Summary of Thermal Transport Properties of 2D LHPs at Room Temperature

|  | n=1 C8:PbBr | n=1 C7:PbBr | n=1 C6:PbBr | n=1 C5:PbBr | n=1 C4:PbBr | n=2 C4:MAPbBr | n=3 C4:MAPbBr |
| --- | --- | --- | --- | --- | --- | --- | --- |
| Thermal Conductivity (W/m·K) | 0.51 ± 0.13 | 0.40 ± 0.12 | 0.29 ± 0.1 | 0.29 ± 0.1 | 0.19 ± 0.07 | 0.27 ± 0.07 | 0.28 ± 0.08 |
| Heat Capacity (J/g·K) | 1.03 ± 0.3 | 0.85 ± 0.26 | 0.72 ± 0.22 | 0.72 ± 0.22 | 0.52 ± 0.16 | 0.63 ± 0.19 | 0.39 ± 0.12 |

**Conclusions:**

In summary, we have measured the thermal conductivity of a series of alkylammonium lead bromide 2D LHPs. We have observed ultralow thermal conductivities ranging from 0.18 – 0.51 W/mK, revealing minimal thermal conductivity suppression relative to bulk MAPbBr perovskites. The thermal conductivity was found to increase monotonically as a function of both the octahedral layer thickness and the organic chain length. Using a thermal conductivity model derived using the Boltzmann transport equation for multilayer structures that exhibit ballistic transport with diffuse interface scattering to understand the origins of these trends, we suggest that thermal transport in the 2D bromide perovskites studied is determined by interface scattering and ballistic transport across the organic subphase rather than phonon scattering within either bulk phase. The efficacy of this model suggests that the unexpectedly high thermal conductivities of 2D bromide perovskites relative to other hybrids and nanomaterials stem from several unique facets of this material. First, the short MFPs of bulk MAPbBr$_3$ perovskites allow for atomically thin inorganic layers that do not exhibit significant suppression of phonon transport. Second, the use of molecular junctions as organic spacers incorporates a subphase that, owing to the unique ballistic transport across these junctions, is actually *more* conductive than the perovskite layer. Thus, 2D LHPs discard the phonon-scattering pathways that limit bulk perovskites in exchange for interface scattering that can be dexterously controlled via the chemistry of the organic subphase. As a result, 2D LHPs can in principle support thermal conductivities an order of magnitude higher than bulk perovskites. Or, since they may be designed to have bulk-like thermal properties as nanocomposites where advantageous and deviant behavior where bulk properties would be deleterious, these materials can be engineered to have even lower thermal conductivities (as for thermoelectric devices). Most critically, since the thermal transport in these materials is driven by ballistic phonon conduction across the organic subphase, the inorganic subphase, and thus the prodigious optoelectronic properties of 2D LHPs, will be largely uncoupled from these pursuits in thermal engineering.

**Methods:**

*Synthesis of 2D LHP crystals.* 2D lead halide perovskite (2D LHP) crystals were made *via* a previously reported cooling-induced crystallization method.[55] Briefly, a solution of $PbBr_2$ was prepared by dissolving PbO (99.9+%, (trace metal basis) <10 microns, powder, ACROS Organic) in concentrated aqueous HBr solution (ACS reagent, 48%, MilliporeSigma) under reflux at 130 °C for 15 minutes. The solution was then allowed to cool to room temperature before a small volume of organic spacer (alkaneamine, L) was added and a white precipitate of $n = 1$ L:PbBr formed. For the syntheses of $n = 1$ 2D LHPs, this solution containing the white precipitate of $n = 1$ L:PbBr was heated on a hot plate set at 130 °C until clear. After that, the clear solution was allowed to cool slowly inside a thermos filled with hot sand at 110 °C to induce crystallization. After a day, crystals of bromide 2D LHPs were collected by suction filtration and dried under reduced pressure for at least 12 hours. For the syntheses of $n = 2$ and 3 2D LHPs, the solution containing the white precipitate of $n = 1$ L:PbBr was mixed with a solution of MABr in concentrated aqueous HBr before the final heating step. Further information on the synthesis and confirmatory material characterization is available in the supporting information.

*Frequency Domain Thermoreflectance (FDTR).* Two continuous wave lasers (Coherent Sapphire 200 CDRH) are used to heat the sample and measure its thermal response: a 488nm pump (which is intensity-modulated via a Conoptics electro-optic modulator) and a 532nm (probe). The former periodically heats the gold surface while the latter continuously monitors the resultant surface temperature modulation at the surface of the transducer via the thermoreflectance of the gold film. The pump is modulated sinusoidally from 100 kHz to 1 MHz. The thermal response is measured using a lock-in amplifier (Zurich Instruments HF2LI), which records the relative phase lag of the measured probe (temperature-induced reflectivity) oscillation as compared to the pump (heat flux) phase. This frequency response is related to the thermal properties of the sample, and the frequency-domain data were fit using a heat transfer model to determine the effective thermal conductivity of the nanostructured macroscopic crystal.[51,56] Laser spot sizes were imaged using a CCD camera prior to measurement. Measurements were done at several locations to ensure no variation in the signal, but since the spot size varies from acquisition to acquisition only one scan was used for the fitting. The resulting thermal conductivity has an uncertainty of ~10% due to input parameter uncertainties (chiefly the spot size of the lasers). Further details on the experimental method employed are available in the supporting information.

*Differential Scanning Calorimetry (DSC)*. DSC was implemented using a TA instruments Discovery DSC. Scans were run using ~5-10 mg of material sealed in Tzero aluminum pans. Heating and cooling scans were run at 10 °C/min ramp/cool rates between 25-60 °C, with 1 minute isothermals in between scans. The heat capacity of a given sample was measured using a reference material (bulk $MAPbBr_3$) with a known heat capacity. Uncertainties reported correspond to the instrument uncertainty as determined by measurement of a second standard (fused silica) using the same measurement scheme. Scan to scan uncertainty was negligible. The lever rule – ($C_{2DLHP} = fC_{MAPbBr,bulk} + (1-f)C_{Alkane,bulk}$), where $f$ is the mass fraction of the perovskite subphase – was used to estimate composite heat capacities using previously measured bulk $MAPbBr_3$ heat capacities and bulk liquid alkane heat capacities.[62] Relative mass fractions were determined using the chemical formula of the nanomaterials.

**Online Supporting Information:**
Supporting information available online: further details of 2D LHP synthesis; characterization by photoluminescence and powder X-ray diffraction (XRD); description and characterization of FDTR instrument, data acquisition, and analysis.

**Acknowledgements:**

We are grateful to Sam Huberman and Liza Lee for assistance with implementation of the FDTR computational model. We thank Sam Winslow for assistance with DSC measurements, and Mahesh Gangishetty and Dan Congreve for use of their thermal evaporation chamber. Synthesis of 2D perovskite materials was supported by the U.S. Department of Energy, Office of Science, Basic Energy Sciences under award number DE-SC0019345. Thermal transport measurements were supported by the U.S. National Science Foundation under award 1452857. N.D. was supported by the MIT Energy Initiative Society of Energy Fellows. Profilometry and powder XRD measurements were performed at the MRSEC Shared Experimental Facilities at MIT, supported by the National Science Foundation under award number DMR-1419807. DSC measurements were performed at the MIT Institute for Soldier Nanotechnologies. The ISN is supported and administered for the US Army by the US Army Research Office (ARO), a part of the Army Research Laboratory (ARL), which is itself a part of the US Army Research Development and Engineering Command (RDECOM). The ISN also receives guidance and oversight from the Assistant Secretary of the Army for Acquisition, Logistics, and Technology (ASA(ALT)) through the Deputy Assistant Secretary of the Army for Research and Technology (DASA(R&T)).

# Supplementary Information for:

## Efficient Thermal Transport across Molecular Chains in Hybrid 2D Lead Bromide Perovskites

Nabeel S. Dahod[1], Watcharaphol Paritmongkol[1,2], William A. Tisdale[1]*

[1]Department of Chemical Engineering, Massachusetts Institute of Technology, Cambridge, Massachusetts, 02139, USA.

[2]Department of Chemistry, Massachusetts Institute of Technology, Cambridge, Massachusetts 02139, USA.

*Correspondence to: tisdale@mit.edu

Contents

1. PbBr 2D LHP synthesis
2. FDTR experimental detail
3. Estimation of material density
4. Parameters used for composite models
5. Alternative models for n-series conductivity trends

# 1. PbBr 2D LHP Synthesis & Preliminary Characterization

*i. PbBr 2D LHP Synthesis*

Crystals of bromide 2D LHPs were synthesized using a previously published cooling-induced crystallization method[1]. Briefly, a solution of $PbBr_2$ was prepared by dissolving PbO (99.9+%, (trace metal basis) <10 microns, powder, ACROS Organic) in concentrated aqueous HBr solution (ACS reagent, 48%, MilliporeSigma) under reflux at 130 °C for 15 minutes. The solution was then allowed to cool to room temperature before a small volume of organic spacer (alkaneamine, L) was added and a white precipitate of $n = 1$ L:PbBr formed. For the syntheses of $n = 1$ 2D LHPs, this solution containing the white precipitate of $n = 1$ L:PbBr was heated on a hot plate set at 130 °C until clear. After that, the clear solution was allowed to cool slowly inside a thermos filled with hot sand at 110 °C to induce crystallization. After a day, crystals of bromide 2D LHPs were collected by suction filtration and dried under reduced pressure for at least 12 hours. For the syntheses of $n = 2$ and 3 2D LHPs, the solution containing the white precipitate of $n = 1$ L:PbBr was mixed with a solution of MABr in concentrated aqueous HBr before the final heating step.

*ii. Photoluminescence Measurements*

In order to confirm that the crystals grown were indeed composed of periodic atomically thin layers of perovskite and organic molecular junctions, photoluminescence spectroscopy was utilized to confirm appropriate quantum confinement given the perovskite quantum well thickness via appearance of an excitonic emission peak (Figure S1). For n=1 PbBr 2D LHPs, this peak is located at ~410 nm, whereas for n=2,3 PbBr 2D LHPs the thicker perovskite layers correspond to weaker quantum confinement and excitonic emission at ~445 nm and ~465 nm, respectively. The inflated linewidth of the n=1 2D LHPs in Figure S1b are likely due to reabsorption of the excitonic emission. The re-absorbed light is emitted at lower energies (higher wavelengths), leading to the broader observed linewidth of the 2D LHPs. This is facilitated by the very small Stokes shift in n=1 2D LHPs and has been reported on elsewhere.[2]

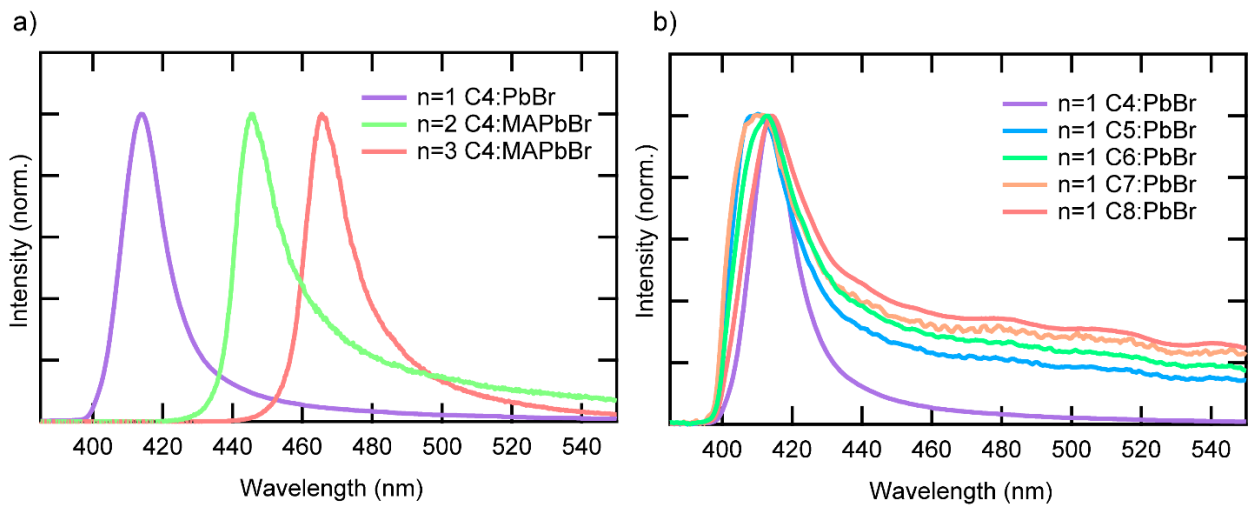

**Figure S1**. Summary of representative PL for PbBr 2D LHPs. a) PL for 2D LHPs at various inorganic sheet (octahedral layer) thicknesses. b) PL for n= 1 2D LHPs with various chain length organic spacer subphases (Cx alkylammonium spacer ions).

Photoluminescence spectra of bromide 2D LHPs were collected by a home-built photoluminescence set up. The output of a 365 nm fiber-coupled LED (Thorlabs, M365FP1) was used as an excitation source and was cleaned by a 390 nm short-pass filter (Semrock, FF01-390/SP-25) before exciting a sample. The emission from a sample was filtered by a 400 nm long-pass filter (Thorlabs, FEL0400) to remove any residual excitation light before being measured by a fiber-coupled spectrometer (Avantes, AvaSpec-ULS2048XL).

*iii. Powder X-Ray Diffraction Measurements*

The long-range order present in the crystals, as well as the retention of the perovskite unit cell within the inorganic sublayer was confirmed with powder X-ray diffraction. These results are readily compared with those corresponding to solved crystal structures of other Ruddlesden-Popper 2D LHPs in the literature, confirming synthesis of pure material and allowing for analysis of structural parameters such as the material density and the organic/inorganic period thicknesses, both of which are readily obtained from analysis of a proposed crystal structure.[1,3,4] Representative powder XRD patterns are shown below in Figure s2.

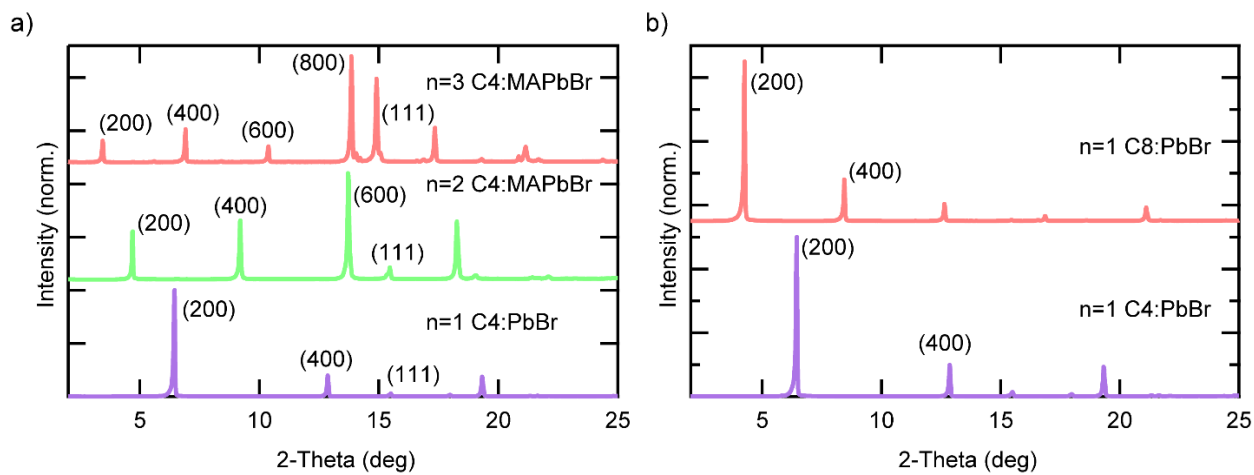

**Figure S2**. Powder X-ray diffraction for structural confirmation. a) Powder XRD for n=1-3 C4:MAPbBr 2D LHPs showing structural confirmation between our materials and those established elsewhere.[1,4] b) Representative powder XRD for n=1 PbBr 2D LHPs with varying organic spacer length, showing structural similarity between our materials and those established elsewhere[1,3]. The consistent decrease in 2-theta for a given peak as a function of increasing spacer length corresponds to the increasing periodicity of the multilayer. The lack of pronounced shift in inorganic (non (h00) planes) suggests the inorganic sublattice is unchanged and well approximated by the bulk unit cell.

Powder X-ray diffraction data was taken using a PANalytical X'Pert Pro MPD X-ray diffractometer (Cu Kα radiation, λ = 1.54184 Å) with High-Speed Bragg-Brentano Optics. A 0.04 rad soller slit, a 1° anti-scatter slit, a 10 mm mask and a programmable divergence slit with an illuminated length of 6 mm were used in the incident beam path. The diffracted beam optics included a 0.04 rad soller slit, a Ni Filter and an automatic receiving slit. The detector was an ultrafast X'Celerator RTMS detector. The angular step in 2θ was 0.04°.

## 2. Frequency Domain Thermoreflectance (FDTR) Experimental Detail

*i. Optical Layout*

The FDTR experimental setup is illustrated schematically in Figure S3. Low noise continuous wave 488 nm and 532 nm diode lasers (both Coherent Sapphire 200 CDRH laser systems) are employed as the pump and probe, respectively. The latter was chosen to optimally leverage the high coefficient of thermoreflectance of the gold transducer at 532 nm. Wavelength-specific optical isolators are used to prevent backscattered light from entering the laser cavities. The intensity modulation of the pump beam required for the experiment is achieved using an electro-optic modulator (EOM), which is driven by an electronic signal amplified from an arbitrary waveform generator. For experimental simplicity and minimal harmonic noise, sine waves with 1 Vpp are fed via the signal generator to the EOM driver. The now-modulated pump beam is then split via a 10:90 non-polarizing cube beam splitter. The lesser portion of the split pump is sent along a path including a micrometer driven translation stage to a photodetector. This signal is sent to the lock-in amplifier as the reference waveform. The larger portion of the split pump is reflected off of a dichroic and sent through the microscope objective to be reflected off of the sample (coated with a Au transducer). The laser spot is spatially Gaussian and oscillates temporally in intensity according to the waveform driving the EOM, successfully implementing the periodic heat flux necessary for the technique.

The probe light is sent through a half waveplate into a polarizing beam splitter (PBS). The reflected light is sent through a quarter waveplate before passing through the dichroic. The use of the PBS and the waveplates allows for efficient separation of the modulated and unmodulated probe light, since light that has reflected off of the sample surface will have a different polarization from that just downstream of the optical isolator. The waveplates can be fine adjusted to optimize reflection of light heading to the sample and signal passed through the PBS post-sample. The laser spots are aligned collinearly following the dichroic, such that the probe monitors the thermoreflectance of roughly the same area that is heated by the pump. After reflection from the sample, the now modulated probe light is collected by the objective and sent back down its original laser path to the PBS. Due to the polarization shift mentioned above the modulated probe light now passes through the PBS before being aligned into another photodetector. The signal measured by this photodetector is sent to the signal input on the lock-in amplifier, which then determines the relative phase lag of the signal modulation from the reference. The micrometer-driven optical delay stage in the 488 nm reference beamline is used to adjust the path length of the reference arm to eliminate absolute phase lag between the reflected probe and reference waveforms.

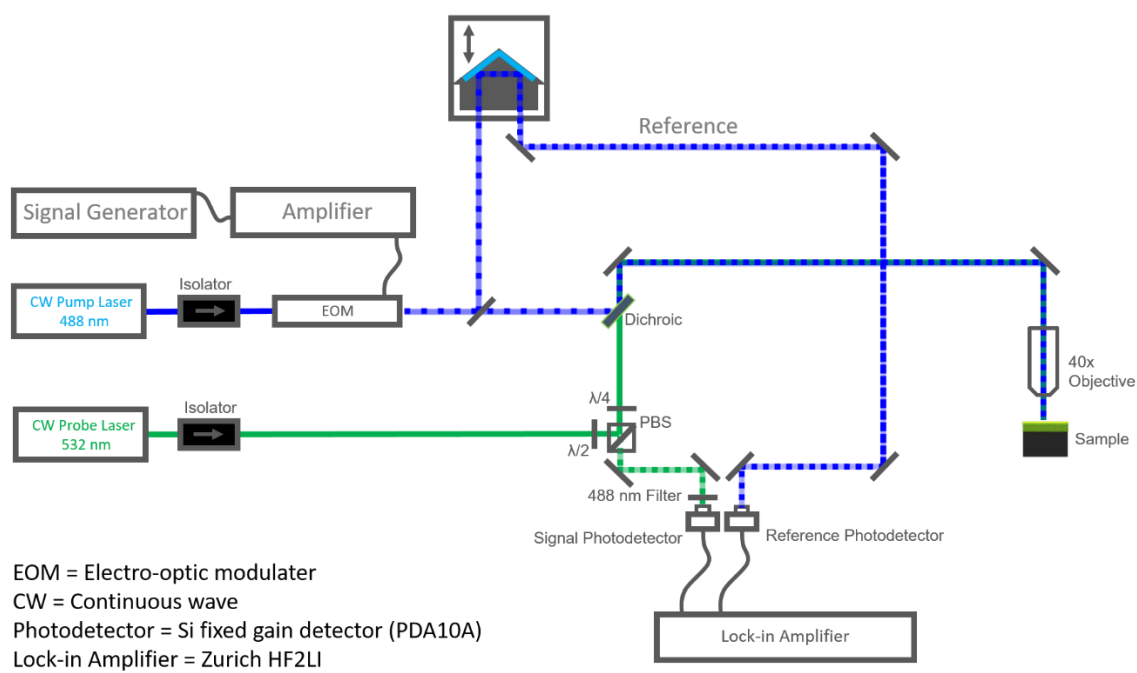

**Figure S3**. Schematic of the optical table layout for FDTR. Colored lines indicate the path of the laser light, with the colors corresponding to those of the pump (blue) and probe (green). When dashed, the lines represent intensity modulated light. Isolators are wavelength specific optical isolators. EOM is a Conoptics model 350 EOM that modulates the intensity of the pump laser light. Amplifier is an analog Conoptics model 25A EOM driver that amplifies a 1 Vpp signal from the signal generator and sends it to the EOM to drive operation. PBS = polarizing beam splitter cube. λ/2 and λ/4 are half and quarter wave plates. 488 nm filter selectively blocks stray pump laser light. Dichroic passes 532 nm laser

light and reflects 488 nm light, 40x objective (Nikon 40x S Plan Fluor) focuses and collects collinear pump/probe laser light to/from sample in backscattering geometry. Si fixed gain detectors measure path-matched reference and signal modulations, which are detected via Lock-in detection and sent to a personal computer via USB.

*ii. Data Acquisition*

Although the operating power for the lasers used is relatively high (200 mW), the surface temperature oscillation required for reliable/physically meaningful data must be relatively small to remain in the linear response regime. Additionally, the steady state temperature rise of the sample should be minimized as much as possible to avoid beam damage and maintain experimental integrity. To this end, a variety of neutral density filters are used to control the power of the reference, pump, and probe beam lines. Typical laser powers (as measured by a power meter at the sample stage) are ~250-300 µW, but higher powers can and typically must be used for more conductive samples (> 1 W/mK).

Similarly, the spot sizes of the pump/probe lasers must be carefully controlled to maximize signal while also probing the dynamics of interest. The spot sizes represent the relevant length scale for the radial component of heat transfer during an FDTR measurement; variation of the spot size can be used to sensitively probe axial, radial, or mixed transport. For instance, if the spot size is larger than the axial penetration depth at a given frequency (as is the case here), then the pump laser can be treated as a 1D plane heating source and the axial component can be ignored. Conversely, when the radial length scale is much smaller than the axial penetration depth, the radial component of heat transport plays a significant role and cannot be ignored. For our experiments, typical spot sizes of ~1.2 µm were found to maximize signal at the above laser powers while remaining within the axial (1D) transport regime. In practice, the spot size must be measured during each individual data acquisition run, which is accomplished through the use of a CCD camera.

The phase response of the sample was measured using a Zurich HF2LI lock-in amplifier. Operation was frequency locked using the output from the reference photodetector, but not phase-locked of course as this is the parameter we expect to monitor. In a typical instance of data acquisition, the phase response is measured by sweeping the modulation frequency fed to the EOM from 100 kHz – 1 MHz and cataloging the measured phase lag at each frequency. This frequency range was chosen due to both the lock-in/EOM operating windows and a relevant span of penetration depths for sensitive thermal property measurement (submicron penetration depths for all perovskite samples).

*iv. Data Fitting and Parameter Extraction*

For consistency with previously reported material thermal conductivities, we utilize the same heat transfer model for analysis posed originally by Cahill, and used ubiquitously since then with respect to FDTR measurements of effective thermal conductivities[5–8]. Many other models have been posed as potentially superior in directly capturing nonequilibrium transport behavior evident in FDTR experiments without the use of an effective thermal conductivity, but such models typically apply to much higher heating frequencies and with respect to far more conductive samples, and so fall outside the scope of this particular report.[9,10] In order to extract an effective thermal conductivity of a macroscopic solid such as a 2D lead halide perovskite crystal, Fourier's law and thus the Cahill model remain sufficient. This model was fit to the experimentally measured phase response (phase lag vs. frequency) using a nonlinear least squares Levenberg-Marquardt algorithm. During fitting, the laser spot size was allowed to vary within experimental error and the thermal conductivity was allowed to vary by an order of magnitude from an initial guess. The uncertainty of the measurement was determined by evaluating the absolute change in measured thermal conductivity for the maximum variance of the spot size within the resolution of the CCD. Typical uncertainties are ~ 10 %, in line with other FDTR measurements previously reported.[11,12] Several representative measurements/fits are shown below in Figures S4.

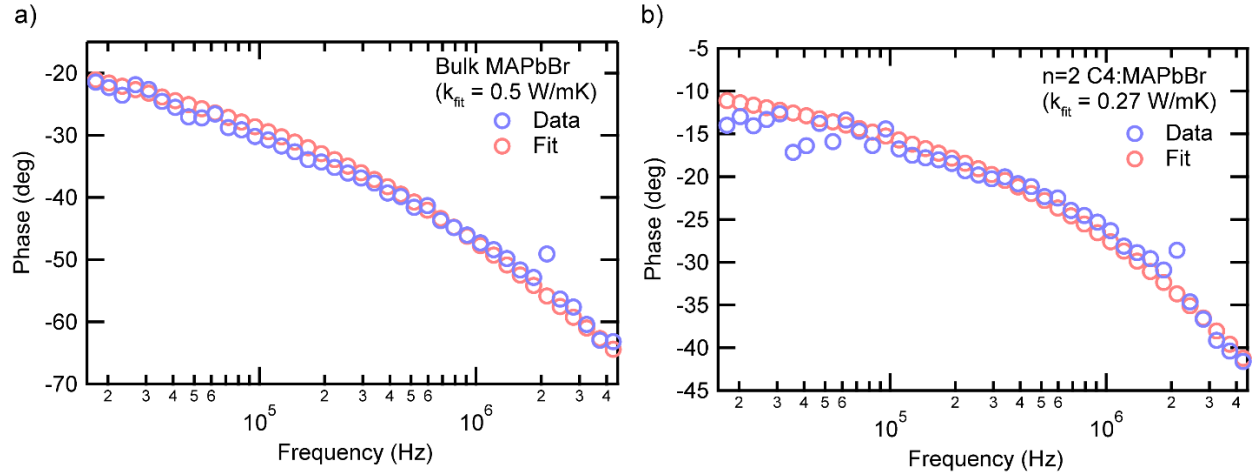

**Figure S4.** Representative data and numerical fits for a single crystal of bulk MAPbBr$_3$ and a 2D LHP crystal. The value for thermal conductivity of bulk MAPbBr$_3$ is in excellent agreement with those previously reported.[12] Note that the heater size (laser spot width) was different for the two data sets shown here.

*iv. Measurement of Relevant System Parameters*

Since an FDTR measurement is sensitive to a variety of system parameters and materials properties, it is important to separately measure all of these parameters except for the parameter of interest, the sample thermal conductivity. With respect to the spot size and transducer thickness/properties, the following approach was used:

To measure the spot size, a CCD camera was used to image the laser spots reflected off of the transducer during a measurement. This image was calibrated using a ruler with 0.1 micrometer gradations, and since the acquired spots are isotropic a line scan through the center of the beam was fit with a Gaussian lineshape to extract the beam waist (representative results shown in Figure S5). Given the high quality of the fit, the primary uncertainty in the spot size comes from the resolution of the image.

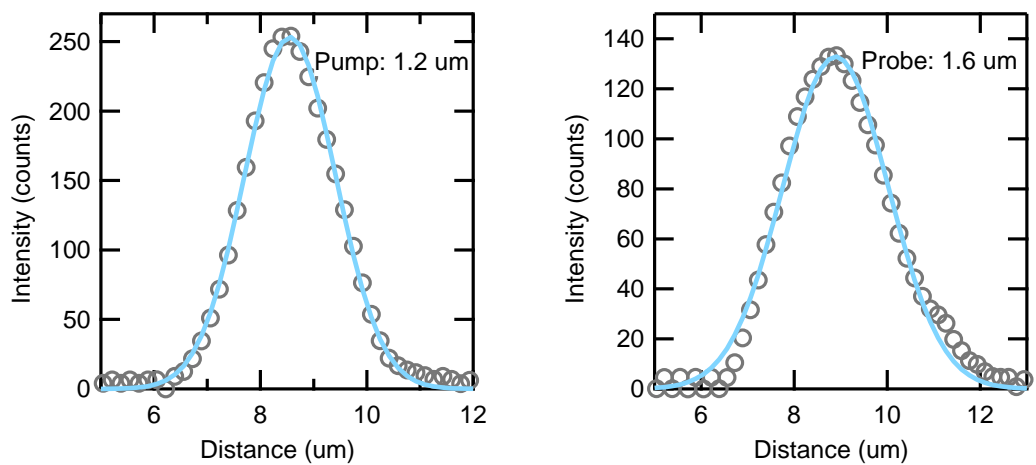

**Figure S5**. Pump (left) and probe (right) laser spots as measured by the CCD (grey markers) and fit by a Gaussian lineshape (blue). Images collected by the CCD are isotropic, thus a line scan is sufficient to measure the spot size within the resolution of the camera. Beam waists shown are the $1/e^2$ beam radius.

The thermal conductivity and volumetric heat capacity of the Au transducer were 314 W/mK and 2500 kJ/m³, as reported elsewhere.[11,12] To measure the thickness of the gold transducer, stylus profilometry was used. First, a reference gold-coated glass film from the same deposition as the samples of interest was scratched using a razor. Then, a Bruker DXT stylus profilometer was used to measure the thickness of the film via a line-scan across this scratch. A representative line-scan is shown in Figure S6.

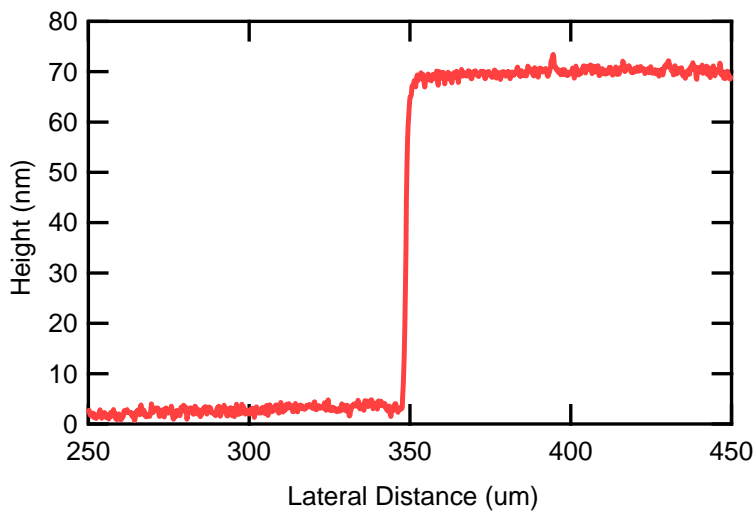

**Figure S6.** A representative profilometry line-scan used to measure the film thickness of a transducer film. The thickness of this Au film was measured to be 67 ±1 nm. Typical films range from 60-150 nm depending on the deposition settings used.

## 3. Estimation of the Mass Density of 2D LHPs

Although differential scanning calorimetry measurements provide reliable measurements of the specific heat capacity, the heat transfer model for FDTR requires knowledge of the volumetric heat capacity (product of the specific heat capacity and the mass density) of a sample. For 2D LHPs, since the generic crystal structure is fairly well established this is simple to estimate even when the precise structural solution is not available. Within the inorganic (perovskite) subphase, the bond lengths and angles are well approximated by that of the pseudocubic bulk unit cell. Thus, the density within this layer is effectively that of the bulk MAPbBr perovskite. This is the case even for n=1 perovskites without a proper methylammonium cation as the spacer binding group is still an ammonium ion. This estimation is consistent with the assumption that the structure/vibrational spectrum of the inorganic subphase is unchanged from that established for bulk MAPbBr. Within the organic subphase, the period thickness is easily extracted by substracting the periodicity as determined via x-ray diffraction measurements by the inorganic subphase thickness. As has been established elsewhere, these thicknesses are consistent across 2D LHPs with different compositions of the inorganic subphase.[1,3] As a result, we can confirm our measurement of the organic subphase with those made for n=1 PbI 2D LHPs precisely measured via single-crystal XRD previously by Billing et al.[13,14]. Knowing the cubic unit cell of the inorganic subphase, octahedral layer thickness, and organic subphase period thickness, the orthorhombic unit cell of PbBr 2D LHPs is readily approximated. Dividing the molecular weight of the 2D LHP as expected from its chemical formula from the volume of this unit cell then directly approximates the mass density of the perovskite. Figure S7 below lists the results of these estimates for the PbBr 2D LHPs studied herein.

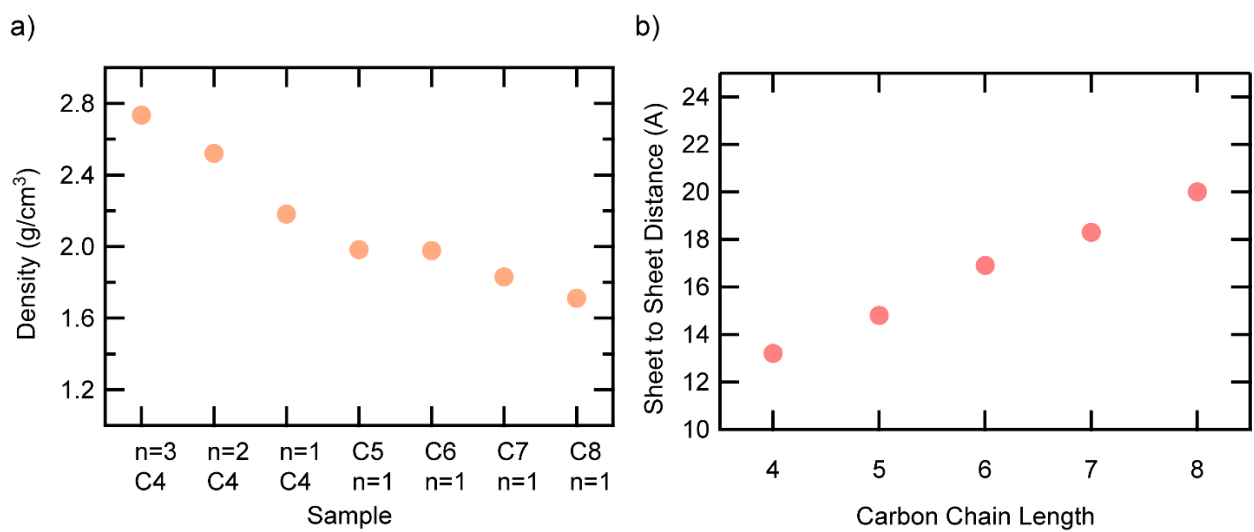

**Figure S7**. a) Estimated mass density of PbBr 2D LHPs. b) Experimentally measured PbBr 2D LHP periodicities

## 4. Parameters Used for Composite Models

In order to determine the thermal conductivity of the PbBr 2D LHPs using the models proposed in the main text, several materials parameters need to be known. Specifically, the period thicknesses, component heat capacities, and component sound velocities are the necessary inputs. The first of these is measured/estimated using XRD as described in the previous section. The heat capacity and sound speed of bulk MAPbBr$_3$ perovskite (380 J/kgK and 1717 m/s) have been measured elsewhere and herein was used for the inorganic subphase.[15] The heat capacity of bulk alkanes were used to approximate the heat capacity within the organic subphase, a reasonable assumption because the organic molecules are known to exhibit significant dynamic disorder (liquid-like motions) at measurement temperatures and have densities comparable to their bulk liquid counterparts.[1,4] The sound speed for ballistic phonon transport across the molecular junctions that comprise the organic subphase has been studied extensively for surface-bound alkyl chains, in systems such as self-assembled monolayers and semiconductor nanocrystals.[16–18] Across all of these measurements, the values for the sound velocity of such phonons are consistent and show essentially linearly proportional sound velocity with increasing alkyl chain length. We use a linear interpolation of the measurements made by Li et al. as an input for calculation of the sound velocity within the organic subphase.[16]

## 5. Alternative models for conductivity trends

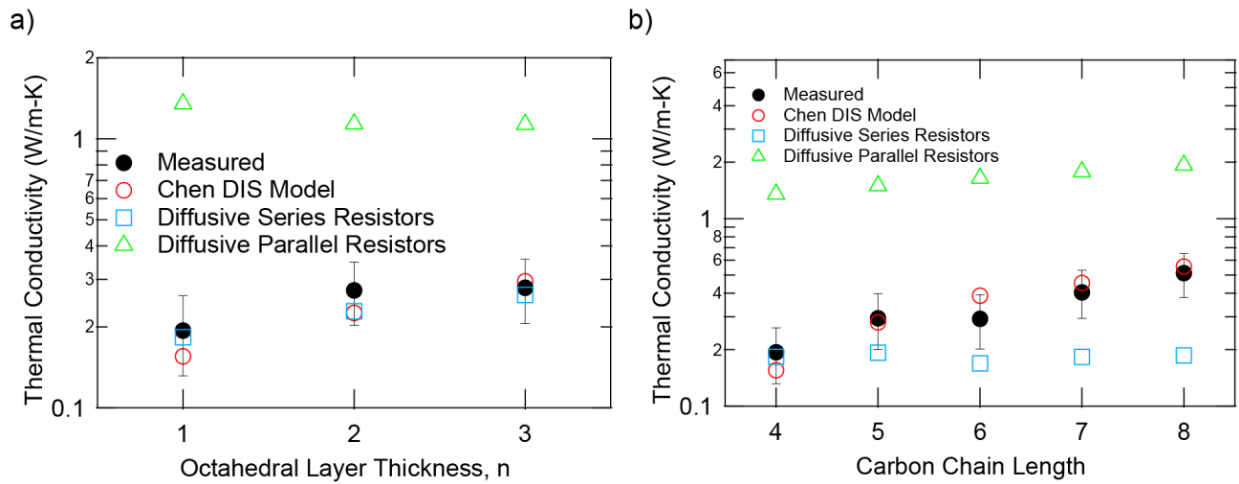

**Figure S8**. Measured thermal conductivity using FDTR for n = 1 BA:PbBr and n = 2,3 BA:MAPbBr 2D LHPs at room temperature (black filled circles) and that expected via the theoretical model in Eq. 1 (red circles), bulk phase thermal resistors in series (blue squares) and parallel (green triangles).